\documentclass[sigconf]{acmart}
\usepackage{subcaption}    % For subfigures
\usepackage{threeparttable}

\AtBeginDocument{%
  }

\acmConference{}{}{}

\begin{document}

% \title{From Plans to Predictions: A Machine Learning Framework for SQL Query Runtime Estimation}

\title{Redefining Cost Estimation in Database Systems: The Role of Execution Plan Features and Machine Learning}

 \author{Utsav Pathak}
 \affiliation{%
   \institution{Dhirubhai Ambani University}
   \city{Gandhinagar}
   \state{Gujarat}
   \country{India}
 }
 \email{202311019@dau.ac.in}

 \author{Amit Mankodi}
 \affiliation{%
   \institution{Dhirubhai Ambani University}
   \city{Gandhinagar}
   \country{India}}
 \email{amit_mankodi@dau.ac.in}

\begin{abstract}
Accurate query runtime prediction is a critical component of effective query optimization in modern database systems. Traditional cost models, such as those used in PostgreSQL, rely on static heuristics that often fail to reflect actual query performance under complex and evolving workloads. This remains an active area of research \cite{Zhu2024,Li2024}], with recent work exploring machine learning techniques to replace or augment traditional cost estimators  \cite{DeepCostModel,BigDataCost}. In this paper, we present a machine learning-based framework for predicting SQL query runtimes using execution plan features extracted from PostgreSQL  \cite{ExplainAnalyze}. Our approach integrates scalar and structural features from execution plans and semantic representations of SQL queries to train predictive models. We construct an automated pipeline for data collection and feature extraction using parameterized TPC-H queries \cite{TPCHBenchmark}, enabling systematic evaluation of multiple modeling techniques. Unlike prior efforts that focus either on cardinality estimation or on synthetic cost metrics, we model the actual runtimes using fine-grained plan statistics and query embeddings derived from execution traces, to improve the model accuracy \cite{QueryOptimizerSurvey}.

We compare baseline regressors, a refined XGBoost model, and a sequential LSTM-based model to assess their effectiveness in runtime prediction. Our dataset includes over 1000 queries generated from TPC-H query templates executed in PostgreSQL with EXPLAIN ANALYZE. Experimental results show that the XGBoost model significantly outperforms others, achieving a mean squared error of 0.3002 and prediction accuracy within ±10\% of the true runtime in over 65\% of cases. The findings highlight the potential of tree-based learning combined with execution plan features for improving cost estimation in query optimizers. Our work contributes a reproducible, extensible framework and provides insights into model selection for integrating learned predictions into real-world database systems.
\end{abstract}





%\received{24 July 2025}
%\renewcommand\footnotetextcopyrightpermission[1]{}

\maketitle

\section{Introduction}

Modern data-driven applications—from business analytics and finance to e-commerce and cloud computing—depend heavily on efficient database systems to ensure responsiveness and scalability \cite{query_lifecycle}. At the core of this efficiency lies the database query optimizer, which is responsible for generating execution plans that minimize query runtime \cite{PostgresDesign}. A key component of this optimization is the cost model, which estimates the execution time and resource usage of alternative plans. However, traditional cost models, such as those used in PostgreSQL \cite{PostgreSQLDatabase}, are heuristic-driven and often inaccurate, especially when applied to complex queries, skewed data distributions, or evolving workloads \cite{QueryOptimizerSurvey}.

These inaccuracies lead to suboptimal plan choices, resulting in increased latency, poor resource utilization, and missed service-level objectives. To address these limitations, recent research has proposed machine learning (ML)-based techniques that learn from past query executions and predict query performance more accurately than static estimators \cite{StatisticalLearning,BigDataCost}. While cardinality estimation has seen significant progress using ML \cite{CardinalityEstimation,AutoCE}, relatively fewer works focus on direct query runtime prediction. Some models estimate runtime based on simplified or synthetic cost functions \cite{DeepCostModel}, while others rely solely on cardinality as a proxy. These approaches often ignore structural details of the execution plan or fail to incorporate real execution statistics—issues we revisit in Section~\ref{sec:related_work}.

In this paper, we present a data-driven approach for predicting SQL query runtimes in PostgreSQL using machine learning models trained on execution plan features \cite{ExplainAnalyze}. In contrast to prior work that often targets cardinality estimation or uses handcrafted operator features \cite{MagpieOptimizer}, our framework directly models total query runtime using real execution traces from PostgreSQL's \texttt{EXPLAIN ANALYZE} \cite{ExplainDocs}. We integrate scalar features (e.g., rows, cost, and execution time), structural plan properties (e.g., node depth, operator hierarchy), and semantic representations of SQL queries (e.g., TF-IDF embeddings \cite{Jones1972}) to enhance model expressiveness.

Our contributions are as follows:

\begin{itemize}
    \item We develop an automated system that collects a dataset of over 1000 TPC-H queries and their corresponding execution plans and actual runtimes from PostgreSQL using \texttt{EXPLAIN ANALYZE} \cite{TPCHBenchmark,TPCHGitHub}, covering variations across all 22 TPC-H query templates \cite{TPCHSpec}.
    
    \item We extract and engineer a diverse set of plan-level features, including more than 20 scalar metrics (e.g., estimated/actual rows, total cost, execution time), 6 structural indicators (e.g., depth, parent-child links), and TF-IDF-based semantic vectors representing the SQL query text \cite{Jones1972}.
    
    \item We evaluate three modeling approaches: (a) baseline regressors (linear \cite{RegressionAnalysis}, SVR \cite{SVR}, random forest \cite{RandomForests}), (b) a refined XGBoost model \cite{XGBoost} using structural and semantic plan features, and (c) a replicated sequential LSTM model \cite{LSTM} inspired by RAAL that treats plan nodes as temporal sequences.
    
    \item We conduct a detailed performance analysis of each model and find that the XGBoost model \cite{XGBoost} outperforms others, achieving an MSE of 0.3002 and ±10\% prediction accuracy in over 65\% of cases. The LSTM model \cite{LSTM} underperforms due to limited data and architectural constraints.
\end{itemize}

The rest of this paper is organized as follows: Section~\ref{sec:background} reviews the limitations of traditional cost models and motivates our approach. Section~\ref{sec:system} describes the system architecture. Section~\ref{sec:features} outlines our feature engineering strategy. Section~\ref{sec:models} presents the modeling approaches used. Section~\ref{sec:experimental_results} details experimental results. Section~\ref{sec:discussion} offers insights and implications. Section~\ref{sec:related_work} discusses related work. Finally, Section~\ref{sec:conclusion} concludes the paper with directions for future research.

\section{Background and Motivation}
\label{sec:background}

Accurate query cost estimation is fundamental to database performance optimization \cite{query_lifecycle}. Traditional cost models, such as those in PostgreSQL \cite{PostgreSQLDatabase,PostgresDesign}, use fixed heuristics and precomputed statistics to estimate the cost of query plans. These methods, while computationally lightweight, often yield inaccurate estimates—particularly in the presence of complex joins, skewed or correlated data, and modern, evolving workloads \cite{QueryOptimizerSurvey}. PostgreSQL’s cost-based optimizer, for instance, assumes uniform data distributions and independence among attributes, which rarely hold in practice \cite{ExplainDocs}. This can mislead the query planner into choosing suboptimal plans, resulting in longer runtimes, inefficient resource usage, and SLA violations in latency-sensitive applications. Although recent research has explored ML-based alternatives \cite{StatisticalLearning,BigDataCost}, many focus primarily on cardinality estimation \cite{CardinalityEstimation,AutoCE} or optimizer feedback \cite{MagpieOptimizer}, with fewer studies directly targeting query runtime prediction using real execution plan features \cite{DeepCostModel}.

In this work, we pose the problem of learned query runtime prediction in PostgreSQL as a supervised learning task over real execution traces \cite{ExplainAnalyze}, with the goal of producing accurate, generalizable, and interpretable models. Our motivation is grounded in the following research questions:

\begin{itemize}
    \item \textbf{RQ1: Can real execution traces from PostgreSQL be effectively leveraged to learn predictive models for total query runtime?} \\
    PostgreSQL's \texttt{EXPLAIN ANALYZE} output provides rich operator-level information, including actual runtimes, estimated and actual rows, and plan hierarchies \cite{ExplainAnalyze,DepeszAnalyzer}. We hypothesize that these statistics can serve as reliable supervision signals for runtime learning.

    \item \textbf{RQ2: How do different categories of features—scalar, structural, and semantic—contribute to the accuracy of learned cost models?} \\
    Execution plans form hierarchical trees of operators. Capturing scalar metrics alone may miss critical structural or contextual patterns \cite{DaliboVisualizer}. We explore whether incorporating these multi-level features improves prediction quality.

    \item \textbf{RQ3: Which types of ML models are most effective for learned runtime prediction in low-data regimes?} \\
    While deep models such as LSTMs \cite{LSTM} offer expressive power, they often require substantial training data. We examine how well simpler models like tree-based regressors \cite{RandomForests,XGBoost} perform relative to deep models under practical dataset sizes.

    % \item \textbf{RQ4: Can benchmark-driven evaluation using TPC-H provide robust, repeatable intersights into learned cost modeling approaches?} \\
    % TPC-H provides a standard suite of parameterized queries over a normalized schema. We use it to construct a controlled yet realistic dataset for evaluating learned runtime models across varied query patterns \cite{TPCHBenchmark,TPCHSpec}.
    \item \textbf{RQ4: Can semantic understanding the queries improve the accurancy of the prediction model?} \\
    By introducing semantic understanding features from query text \cite{Jones1972,SentenceBERT} in addition to execution plan features, we evaluate the performance of the prediction model.
\end{itemize}

These questions form the foundation for our system design and evaluation, motivating the architecture we present in the next section.

\section{Our Methodology}
\label{sec:system}
\begin{figure}[t]
  \centering
  \includegraphics[width=0.4\textwidth]{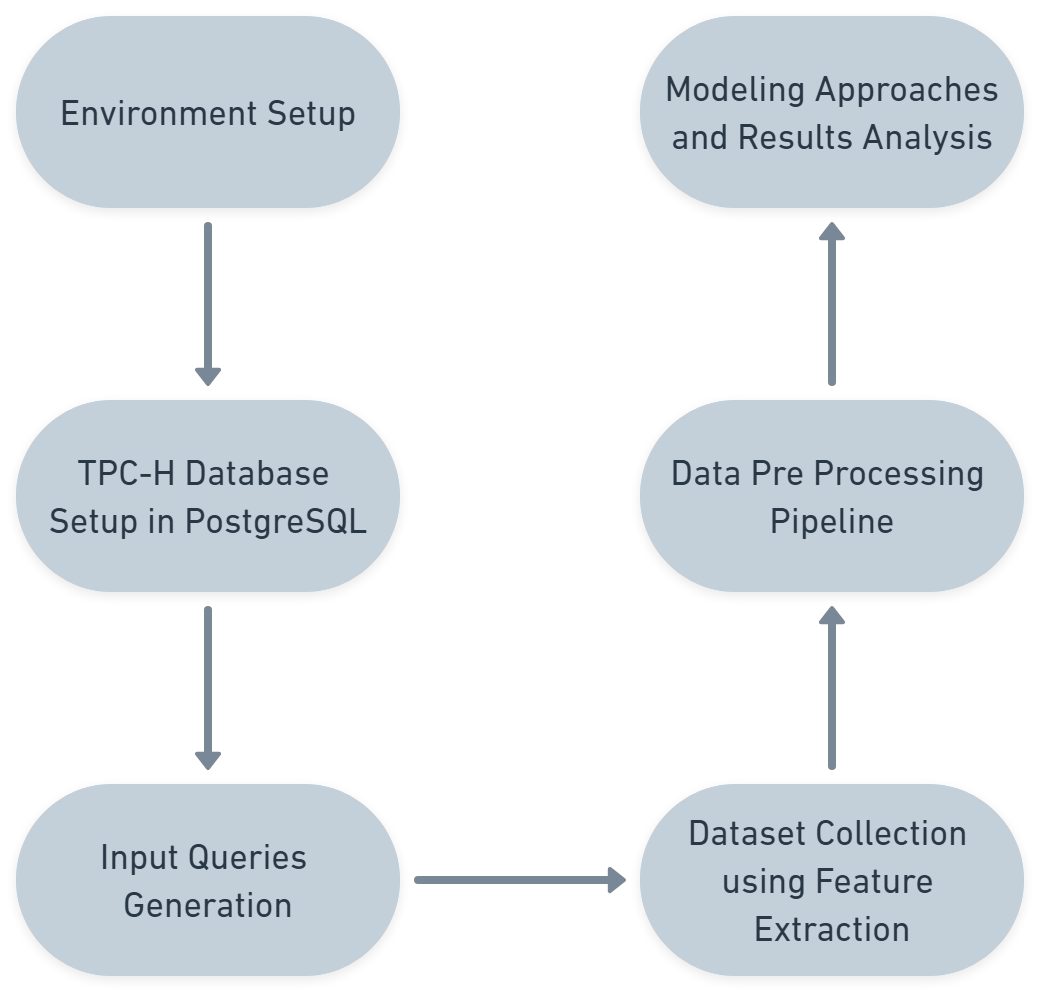}
  \caption{Overview of Our Methodology}
  \label{fig:flow}
\end{figure}
This section describes the architecture of our end-to-end framework for learning query runtime prediction models using PostgreSQL \cite{PostgreSQLDatabase} and the TPC-H benchmark \cite{TPCHBenchmark}. The system automates the entire pipeline—from query generation and execution to feature extraction and model training—enabling systematic experimentation and evaluation. It is organized into five key stages: query generation, query execution and feature extraction, data preprocessing, and model training. Figure \ref{fig:flow} shows the overview of our methodology.

\subsection{Query Generation}
We construct our workload using the TPC-H benchmark, which consists of 22 analytical query templates designed for decision support \cite{TPCHSpec,TPCHGitHub}. To generate a diverse set of queries, we systematically vary template parameters such as date ranges, region filters, and numeric thresholds. This produces a broad spectrum of queries that differ semantically while preserving structural patterns, resulting in over 1000 unique SQL queries executed against a PostgreSQL instance \cite{PostgreSQLDatabase}.

\subsection{Query Execution and Feature Extraction}
Each query is executed with \texttt{EXPLAIN ANALYZE} enabled to capture real execution plans \cite{ExplainAnalyze,ExplainDocs}. We automate this process via a shell script (\texttt{run\_all.sh}) that logs each query and its corresponding plan output in JSON format. From these plans, we extract structured features across three main categories:

\begin{itemize}
    \item \textbf{Scalar Features:} Estimated and actual row counts, total cost, startup cost, and execution time \cite{DepeszAnalyzer}.
    \item \textbf{Structural Features:} Properties such as plan node type, depth in the tree, and parent-child relationships \cite{DaliboVisualizer}.
    \item \textbf{Semantic Features:} Dense query representations generated using the \texttt{all-MiniLM-L6-v2} transformer model \cite{SentenceBERT}, capturing contextual semantics of the SQL text and enhancing the model’s ability to differentiate between similar plan structures arising from semantically different queries.
\end{itemize}

Each plan is decomposed into its constituent nodes, and the resulting feature vectors are compiled into a structured CSV file, where each row corresponds to a plan node associated with its parent query \cite{CSVFormat}. 

\subsection{Data Preprocessing}
Before model training, we encode categorical fields (e.g., node type) using label encoding or one-hot encoding \cite{ScikitLearn}. Derived features such as cardinality ratios (actual vs. estimated rows) and normalized costs are also computed to improve predictive signal \cite{CardinalityEstimation}. Missing values are handled using filtering or imputation strategies. Semantic features are generated using TF-IDF over the corpus of SQL queries \cite{Jones1972}, enabling integration into hybrid learning models.

\subsection{Modeling Pipeline}
The final stage involves training machine learning models on the processed dataset \cite{StatisticalLearning}. We implement several model families:

\begin{itemize}
    \item \textbf{Baseline Regressors:} Including linear regression \cite{RegressionAnalysis}, support vector regression (SVR) \cite{SVR}, and random forest regressors \cite{RandomForests}.
    \item \textbf{XGBoost Model:} A refined gradient-boosted tree ensemble that incorporates both structural and semantic features \cite{XGBoost}.
    \item \textbf{Sequential LSTM Model:} A simplified implementation of RAAL, treating the plan as a sequence of node vectors to learn temporal dependencies \cite{LSTM}.
\end{itemize}

Each model is trained to predict the total query runtime given node- and query-level features. The modular design of our system allows controlled experimentation with alternative features and architectures, supporting reproducibility and extensibility for future research.

\section{Feature Engineering}
\label{sec:features}

Accurate query runtime prediction relies heavily on the expressiveness and quality of the features extracted from execution plans \cite{ExplainAnalyze}. Our feature engineering process is designed to capture key signals from both the raw execution metrics and the structural and semantic context of each query. We organize features into three primary categories: scalar, structural, and semantic, as summarized in Table~\ref{tab:query_features}. Table \ref{tab:csv} show our dataset for a single query extracted from our TPC-H Database.

\begin{table}[t]
  \caption{Query Execution Plan Features}
  \label{tab:query_features}
  \centering
  \begin{tabular}{cll}
    \toprule
    \textbf{Sr. No.} & \textbf{Feature Name} & \textbf{Type/Units} \\
    \midrule
    1 & Original Query         & Text \\
    2 & Node Type             & Categorical \\
    3 & Parallel Aware        & Boolean \\
    4 & Startup Cost          & Numeric (cost units) \\
    5 & Total Cost            & Numeric (cost units) \\
    6 & Plan Rows             & Numeric (rows) \\
    7 & Plan Width            & Numeric (bytes) \\
    8 & Actual Startup Time   & Numeric (ms) \\
    9 & Actual Total Time     & Numeric (ms) \\
    10 & Actual Rows          & Numeric (rows) \\
    11 & Loops                & Numeric (count) \\
    12 & Base Cardinality     & Numeric (rows) \\
    13 & Output Cardinality   & Numeric (rows) \\
    14 & Input Cardinality    & Numeric (rows) \\
    \bottomrule
  \end{tabular}
\end{table}

\subsection{Scalar Features}
Scalar features capture node-level numeric information as reported by PostgreSQL's \texttt{EXPLAIN ANALYZE} \cite{ExplainAnalyze,DepeszAnalyzer}. These include:

\begin{itemize}
    \item \textbf{Estimated Metrics:} Startup cost, total cost, estimated number of rows, and plan width.
    \item \textbf{Actual Execution Metrics:} Actual startup time, total execution time, actual row count, and number of loops.
    \item \textbf{Derived Metrics:} Ratios of actual to estimated rows (cardinality error) \cite{CardinalityEstimation}, cost per row, time per loop, and plan-wide statistics such as average depth and total node count.
\end{itemize}

These features form the quantitative foundation for runtime modeling, particularly effective in classical regression setups \cite{RegressionAnalysis}.

\subsection{Structural Features}
To capture hierarchical execution flow, we extract structural features from the plan tree \cite{DaliboVisualizer}:

\begin{itemize}
    \item \textbf{Node Depth:} Position of the node in the tree hierarchy.
    \item \textbf{Parent-Child Relationships:} Useful for modeling the execution order and plan shape.
    \item \textbf{Subtree Cardinality:} Number of descendant nodes, indicating computational depth.
    \item \textbf{Node Type:} Encoded operator labels (e.g., Seq Scan, Hash Join, Nested Loop).
\end{itemize}

These features encode the execution strategy and flow, enabling tree-based models to reason about operator placement and execution dependencies \cite{RandomForests,XGBoost}.

\subsection{Semantic Features}
Beyond plan structure, we capture query-level semantics to embed the high-level intent of the SQL query \cite{Jones1972}:

\begin{itemize}
    \item \textbf{Query text Embeddings:} Each query is encoded using dense semantic embeddings derived from the \texttt{all-MiniLM-L6-v2} transformer model \cite{SentenceBERT}. These representations capture the contextual meaning of the entire SQL query rather than relying on traditional token frequency statistics.

    \item \textbf{Plan-Level Aggregations:} Aggregated representations across plan nodes are used in hybrid models to capture global structure.
\end{itemize}
% Creating the simplified table without MP and MR columns
\begin{table*}[t]
\centering
\caption{CSV Features Collected from the TPC-H Database for a single query}
\label{tab:csv}
\begin{threeparttable}
\begin{tabular}{ccccccccccc}
\hline
\textbf{LN} & \textbf{PL} & \textbf{NT} & \textbf{PA} & \textbf{SC} & \textbf{TC} & \textbf{PW} & \textbf{ST} & \textbf{TT} & \textbf{IC} & \textbf{OC} \\
\hline
1.1 & -1 & Limit & FALSE & 43832.5 & 43833.81 & 16 & 880.635 & 949.711 & 10 & 10 \\
2.1 & 1.1 & Gather & FALSE & 43832.5 & 829129.72 & 16 & 880.632 & 949.706 & 4 & 10 \\
3.1 & 2.1 & Hash Join & TRUE & 42832.5 & 227992.42 & 16 & 847.728 & 847.889 & 2500405 & 4 \\
4.1 & 3.1 & Seq Scan & TRUE & 0 & 138501.72 & 12 & 0.396 & 413.957 & 6001215 & 2000405 \\
4.2 & 3.1 & Hash & TRUE & 32578 & 32578 & 8 & 156.572 & 156.573 & 500000 & 500000 \\
5.1 & 4.2 & Seq Scan & TRUE & 0 & 32578 & 8 & 0.377 & 91.72 & 1500000 & 500000 \\
\hline
\end{tabular}
\begin{tablenotes}[flushleft]
\small
\item \textbf{LN}: Level Number, \textbf{PL}: Parent Level, \textbf{NT}: Node Type, \textbf{PA}: Parallel Aware
\item \textbf{SC}: Startup Cost, \textbf{TC}: Total Cost, \textbf{PW}: Plan Width, \textbf{ST}: Actual Startup Time (ms)
\item \textbf{TT}: Actual Total Time (ms), \textbf{IC}: Input Cardinality, \textbf{OC}: Output Cardinality
\end{tablenotes}
\end{threeparttable}
\end{table*}

Semantic features are particularly beneficial for models like LSTM \cite{LSTM}, which benefit from richer representations that contextualize operator behavior within query semantics.

\subsection{Feature Encoding and Normalization}
Categorical variables (e.g., node type, join strategy) are encoded using one-hot or label encoding depending on model requirements \cite{ScikitLearn}. Numeric features are normalized via z-score standardization or min-max scaling. Sparse or missing values are handled through imputation or filtering based on threshold-based rules. These steps ensure feature consistency and gradient stability during training \cite{StatisticalLearning}.

By combining scalar execution signals, hierarchical structural properties, and SQL-level semantics, our feature engineering approach provides a robust and expressive representation of query plans, which drives the performance of the models discussed in subsequent sections.

\section{Modeling Approaches}
\label{sec:models}

In this section, we present the machine learning models used for query runtime prediction \cite{StatisticalLearning}. Our objective is to learn a mapping from engineered plan features to accurate predictions of total query execution time. We explore a range of models—starting from classical regressors to advanced architectures—that leverage both structural and semantic information.

\subsection{Training and Evaluation}
All models are trained using the same dataset comprising over 1000 TPC-H query instances \cite{TPCHBenchmark,TPCHGitHub}, with an 80-20 train-test split. We evaluate performance using three metrics:

\begin{itemize}
    \item \textbf{Mean Squared Error (MSE):} Captures the average squared difference between predicted and actual runtimes:
    \[
    \text{MSE} = \frac{1}{n} \sum_{i=1}^{n} (y_i - \hat{y}_i)^2
    \]
    where \( y_i \) is the true runtime and \( \hat{y}_i \) is the predicted runtime. MSE penalizes larger errors more heavily \cite{RegressionAnalysis}.

    \item \textbf{R-squared (R²):} Measures the proportion of variance in the actual runtimes explained by the model:
    \[
    R^2 = 1 - \frac{\sum_{i=1}^{n} (y_i - \hat{y}_i)^2}{\sum_{i=1}^{n} (y_i - \bar{y})^2}
    \]
    where \( \bar{y} \) is the mean of actual runtimes. Higher R² indicates better model fit \cite{RegressionAnalysis}.

    \item \textbf{Accuracy within ±10\%:} The percentage of predictions where the relative error is within ±10% of the true runtime:
    \[
    \text{Accuracy}_{10\%} = \frac{1}{n} \sum_{i=1}^{n} \mathbb{I} \left[ \frac{|y_i - \hat{y}_i|}{y_i} \leq 0.1 \right] \times 100
    \]
    This metric offers an interpretable benchmark of model precision in a real-world scenario.
\end{itemize}

These metrics provide a balanced view of model quality—covering both statistical performance (MSE, R²) and practical prediction fidelity (±10\% accuracy). Hyperparameters are tuned using grid search or validation splits \cite{ScikitLearn}.

\subsection{Baseline Models}
We begin with a set of standard regression models that serve as baselines:

\begin{itemize}
    \item \textbf{Linear Regression:} A simple model assuming a linear relationship between input features and runtime. It provides interpretability but often underperforms on non-linear data \cite{RegressionAnalysis}.

    \item \textbf{Support Vector Regression (SVR):} Utilizes a radial basis function kernel to capture non-linear relationships. While effective, SVR can be computationally expensive for large feature sets \cite{SVR}.

    \item \textbf{Random Forest Regression:} An ensemble of decision trees trained on feature subsets. It is robust to overfitting and captures feature interactions, though it may struggle with capturing plan-wide dependencies \cite{RandomForests}.
\end{itemize}

These models provide a useful reference point to evaluate the benefits of more expressive learning architectures.

\subsection{XGBoost with Plan Features and Query Text Embeddings}
We implement a refined model using XGBoost, a tree-based gradient boosting algorithm well-suited for tabular regression problems \cite{XGBoost}. Our XGBoost model integrates three categories of input features:

\begin{itemize}
    \item \textbf{Scalar Plan Features:} Including both estimated (e.g., cost, rows) and actual metrics (e.g., execution time, loops) \cite{ExplainAnalyze}.
    \item \textbf{Structural Features:} Such as node depth, parent-child relationships, and node types \cite{DaliboVisualizer}.
    \item \textbf{Query Text Embeddings:} Sentence-level vector representations derived from SQL text using the \texttt{all-MiniLM-L6-v2} transformer model \cite{SentenceBERT}. These embeddings capture the semantic and syntactic nuances of the query beyond surface-level token patterns \cite{Jones1972}.
\end{itemize}

XGBoost’s ability to handle heterogeneous inputs and model complex interactions makes it ideal for capturing both local and global signals from execution plans.

\subsection{LSTM-Based Sequential Model}
To explore deep learning for plan modeling, we replicate a simplified version of the Resource-Aware Attentional LSTM (RAAL) \cite{DeepCostModel,LSTM}. In our version, each execution plan is flattened into a sequence of node feature vectors, which are passed through a stacked LSTM followed by dense layers to regress total runtime.

Due to limited dataset size and the absence of resource usage tracking (as required by RAAL), we omit the attention mechanism and resource-aware components. Nevertheless, the model serves as a benchmark for evaluating the viability of sequence models in capturing temporal and hierarchical plan patterns \cite{LSTM}.

\section{Experimental Results}
\label{sec:experimental_results}

In this section, we evaluate the performance of the models described in Section~\ref{sec:models} on the dataset generated as described in section \ref{sec:features} for query runtime prediction \cite{StatisticalLearning}. The evaluation emphasizes accuracy, robustness, and model generalization.

\subsection{Environment Setup}
All experiments were conducted on a Linux machine running Ubuntu 20.04 \cite{UbuntuWebsite} with 16 GB RAM, Intel Core i5-1135G7 processor (4 cores, 8 threads), and PostgreSQL version 14.7 \cite{PostgreSQLDatabase}. The dataset, described in Section~\ref{sec:system}, includes over 1000 plan instances generated from parameterized TPC-H query templates \cite{TPCHBenchmark,TPCHGitHub}. Each query was executed using \texttt{EXPLAIN ANALYZE} to collect runtime data and execution plans \cite{ExplainAnalyze}. The complete pipeline—from query execution to plan parsing and feature engineering—is illustrated in Figure~\ref{fig:flow}.

\subsection{Evaluation Metrics}
We evaluate all models using the metrics introduced in Section~\ref{sec:models}: Mean Squared Error (MSE), R-squared (R²), and Accuracy within ±10\% of true runtime \cite{RegressionAnalysis}. These metrics collectively capture both statistical fidelity and practical prediction performance.

All models were trained using an 80-20 train-test split, ensuring stratified distribution of query templates across both sets. Hyperparameters were tuned via grid search or validation splits depending on the model \cite{ScikitLearn}.

\subsection{Results of Baseline Models}
Figures~\ref{fig:combined}(a), \ref{fig:combined}(b) and Table~\ref{tab:baseline_metrics} summarizes the performance of baseline models. As discussed earlier, these models use scalar and basic structural features without explicit semantic context or hierarchical modeling \cite{ExplainAnalyze,DaliboVisualizer}.

% \begin{figure*}[t]
%   \centering
%   \begin{subfigure}[t]{0.30\textwidth}
%     \centering
%     \includegraphics[width=\linewidth]{model_accuracy_bar_chart.png}
%     \caption{Runtime Comparison for LSTM}
%     \label{fig:accuracy}
%   \end{subfigure}
%   \hfill
%   \begin{subfigure}[t]{0.30\textwidth}
%     \centering
%     \includegraphics[width=\linewidth]{mse_comparison_bar_chart.png}
%     \caption{Runtime Comparison for XGBoost}
%     \label{fig:mse}
%   \end{subfigure}
%   \caption{Comparison of Accuracy and MSE of Initial Experiments}
%   \label{fig:initial_side_by_side}
% \end{figure*}

\begin{table}[t]
  \caption{Performance Metrics for Baseline Models}
  \label{tab:baseline_metrics}
  \centering
  \begin{tabular}{lccc}
    \toprule
    \textbf{Model} & \textbf{MSE} & \textbf{R\textsuperscript{2} Score} & \textbf{Acc ($\pm$10\%)} \\
    \midrule
    Linear Regression         & 0.7289 & 0.3385 & 10.15\% \\
    Support Vector Regressor  & 0.4670 & 0.5762 & 28.00\% \\
    Random Forest Regressor   & 0.4185 & 0.6202 & 49.23\% \\
    Feedforward Neural Network & 0.4814 & 0.5631 & 28.92\% \\
    XGBoost (Flat Features)   & 0.3607 & 0.6727 & 57.23\% \\
    \bottomrule
  \end{tabular}
\end{table}

The results show that classical models like linear regression \cite{RegressionAnalysis} and SVR \cite{SVR} are insufficient for capturing the complexity of query plans. Random Forests \cite{RandomForests} and XGBoost with flat features \cite{XGBoost} improve performance, but still lack full contextual modeling.

\subsection{Refined XGBoost Model}
The enhanced XGBoost model, described in Section~\ref{sec:models}, integrates semantic features as query text embeddings \cite{SentenceBERT} to scalar, structural features \cite{ExplainAnalyze,DaliboVisualizer}. It outperforms all other models across evaluation metrics:

% \begin{itemize}
%     \item \textbf{MSE:} 0.3002
%     \item \textbf{R²:} 0.9512
%     \item \textbf{±10\% Accuracy:} 65.52\%
% \end{itemize}

Table \ref{tab:combined_metrics} shows the results of the Refined XGBoost model performance. This performance demonstrates XGBoost’s ability to model non-linear dependencies and exploit hierarchical plan patterns \cite{XGBoost}. Feature importance analysis identified estimated rows, actual rows, and node depth as key predictors \cite{CardinalityEstimation}.

\subsection{LSTM-Based Model}
The LSTM model, using sequential representations of plan nodes \cite{LSTM}, underperformed due to data sparsity and lack of architecture-specific tuning:
% Creating a combined table for both sets of metrics
\begin{table}[t]
\centering
\caption{Performance Metrics for Refined XGBoost and LSTM}
\label{tab:combined_metrics}
\begin{tabular}{lcc}
\toprule
\textbf{Metric} & \textbf{Refined XGBoost} & \textbf{LSTM} \\
\midrule
MSE & 0.3002 & 0.8915 \\
R² & 0.9512 & 0.0216 \\
±10\% Accuracy & 65.52\% & 17.24\% \\
\bottomrule
\end{tabular}
\end{table}

% \begin{itemize}
%     \item \textbf{MSE:} 0.8915
%     \item \textbf{R²:} 0.0216
%     \item \textbf{±10\% Accuracy:} 17.24\%
% \end{itemize}

Table \ref{tab:combined_metrics} shows the results of the LSTM model performance. This result suggests that while sequence models may be effective in theory \cite{DeepCostModel}, they require substantially larger datasets and architectural refinement to compete with tree-based models in this setting.

\subsection{Comparative Analysis}
\begin{figure*}[t]
  \centering
  \begin{subfigure}[t]{0.24\textwidth}
    \centering
    \includegraphics[width=\linewidth]{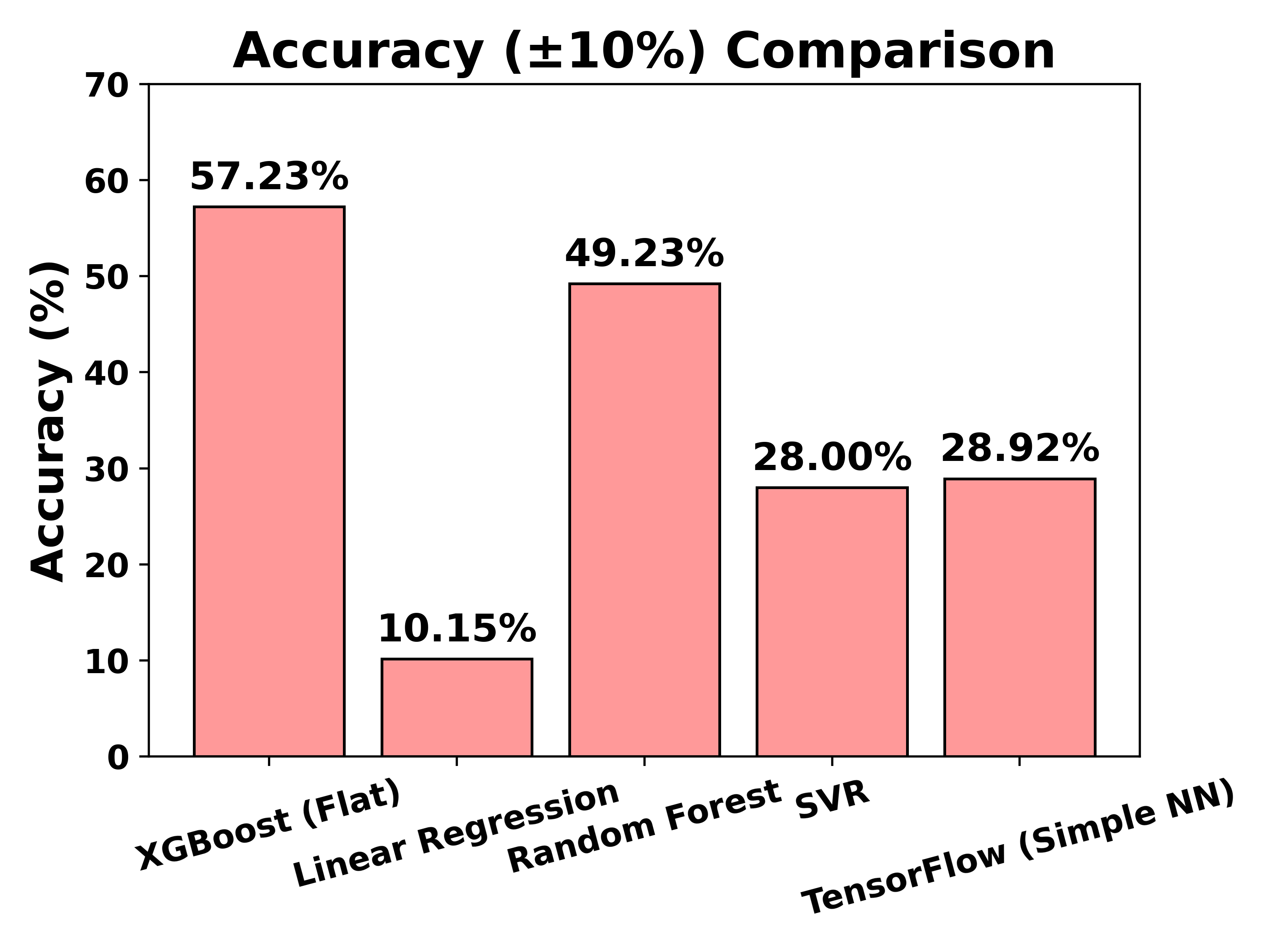}
    \caption{Comparison of Accuracy for the Initial Moels}
    \label{fig:accuracy}
  \end{subfigure}
  \hfill
  \begin{subfigure}[t]{0.24\textwidth}
    \centering
    \includegraphics[width=\linewidth]{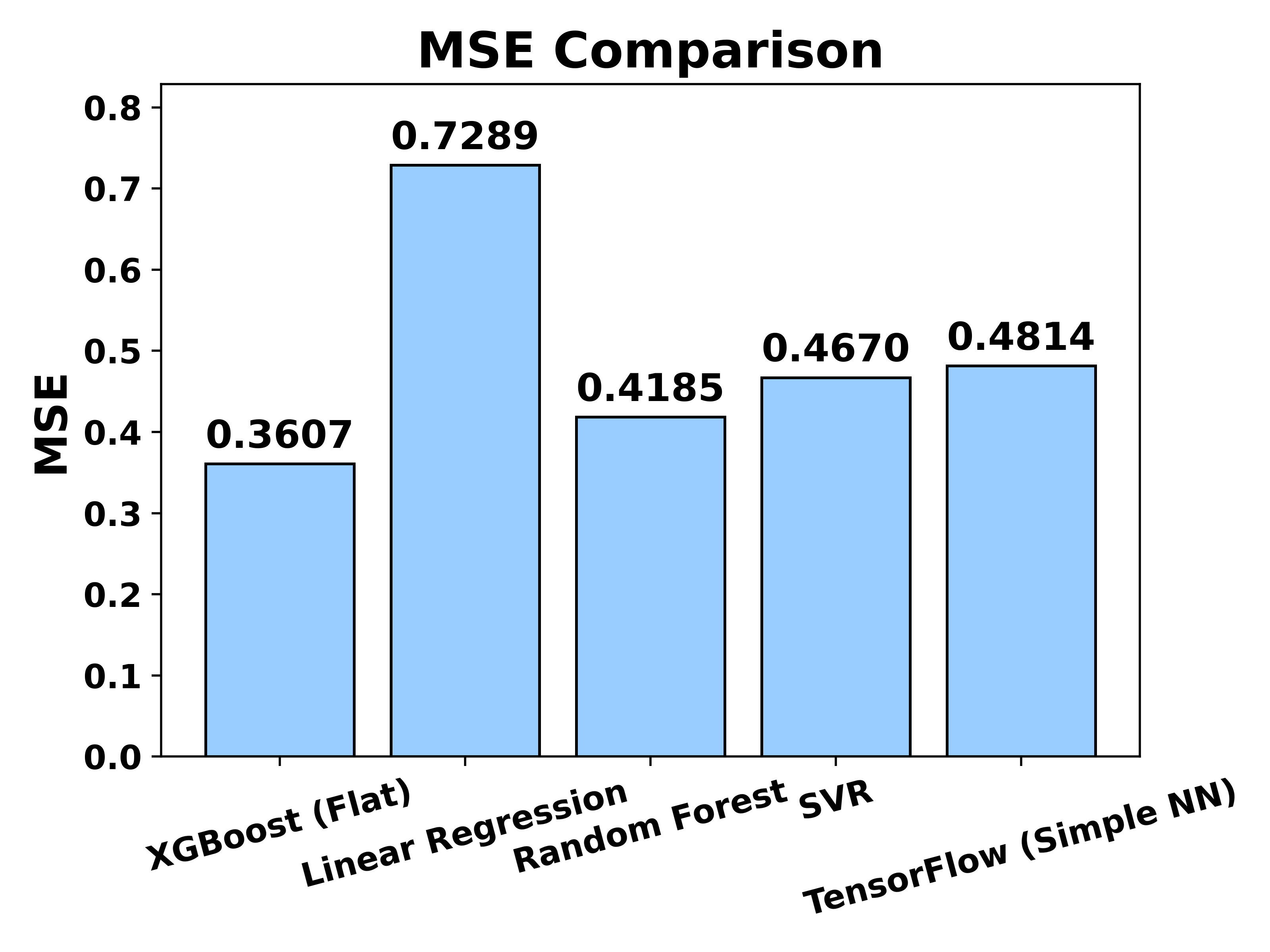}
    \caption{Comparison of MSE Score for the Initial Models}
    \label{fig:mse}
  \end{subfigure}
  \hfill
  \begin{subfigure}[t]{0.24\textwidth}
    \centering
    \includegraphics[width=\linewidth]{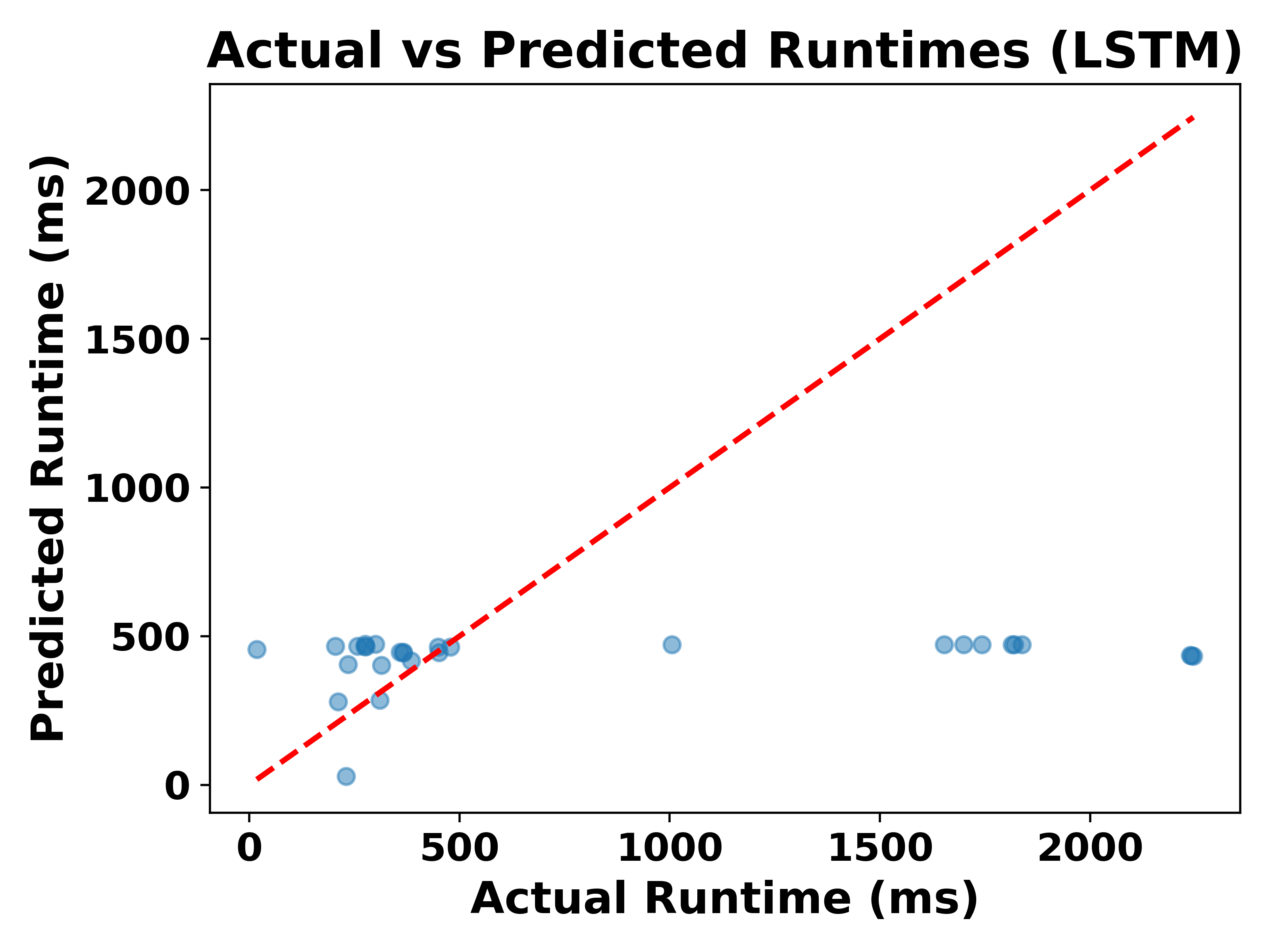}
    \caption{Runtime Comparison for LSTM (Predictions)}
    \label{fig:image1}
  \end{subfigure}
  \hfill
  \begin{subfigure}[t]{0.24\textwidth}
    \centering
    \includegraphics[width=\linewidth]{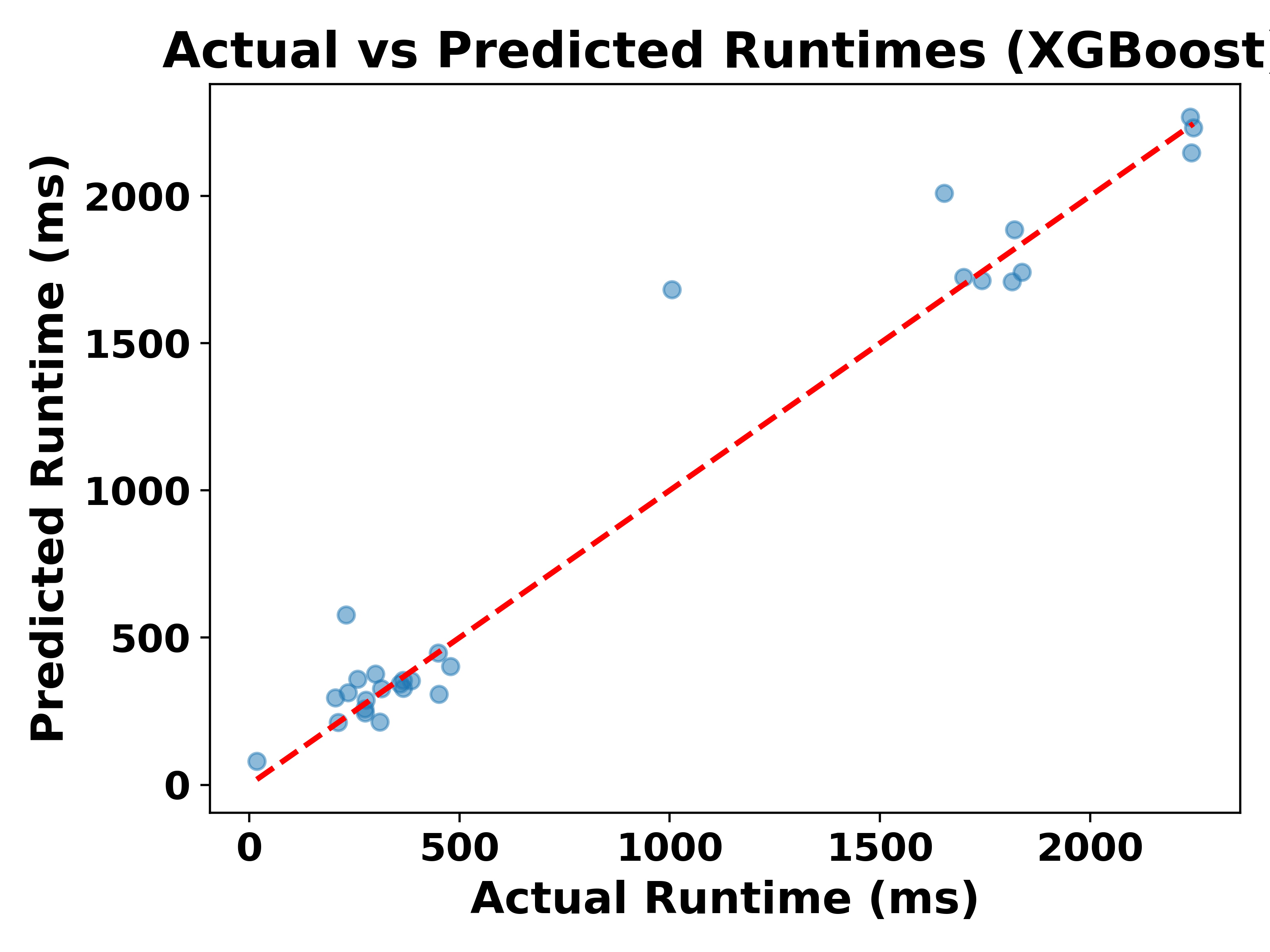}
    \caption{Runtime Comparison for Refined XGBoost (Predictions)}
    \label{fig:image2}
  \end{subfigure}
  \caption{Comparative Analysis of Initial [(a) and (b)] and Refined [(c) and (d)] Experiment Results }
  \label{fig:combined}
\end{figure*}

Figure~\ref{fig:combined}(c) and \ref{fig:combined}(d) presents side-by-side runtime prediction comparisons for XGBoost \cite{XGBoost} and LSTM \cite{LSTM}. As shown, XGBoost achieves tighter prediction bounds and lower residual variance.

% \begin{figure*}[t]
%   \centering
%   \begin{subfigure}[t]{0.40\textwidth}
%     \centering
%     \includegraphics[width=\linewidth]{lstm.png}
%     \caption{Runtime Comparison for LSTM}
%     \label{fig:image1}
%   \end{subfigure}
%   \hfill
%   \begin{subfigure}[t]{0.40\textwidth}
%     \centering
%     \includegraphics[width=\linewidth]{xgboost_predictions_plot.png}
%     \caption{Runtime Comparison for XGBoost}
%     \label{fig:image2}
%   \end{subfigure}
%   \caption{Side-by-side comparison of LSTM and XGBoost prediction performance.}
%   \label{fig:side_by_side}
% \end{figure*}

The findings confirm that models leveraging tree structures and feature hierarchies—like XGBoost \cite{XGBoost}—offer better generalization under data constraints, while sequence models remain promising yet data-hungry \cite{LSTM}.

\subsection{Summary}
These results validate the effectiveness of hybrid tree-based models for query runtime prediction in PostgreSQL \cite{PostgreSQLDatabase}. XGBoost, equipped with scalar, structural, and semantic features \cite{SentenceBERT,ExplainAnalyze}, provides a reliable, interpretable, and accurate estimation mechanism. The LSTM-based model \cite{LSTM} illustrates the challenges of sequence modeling in low-data regimes, setting a foundation for future exploration with larger workloads and richer supervision.

\section{Discussion}
\label{sec:discussion}

This section reflects on the experimental results presented earlier, distills high-level takeaways, and connects them with the core research questions (RQs) posed in Section~\ref{sec:background}. We analyze the relative effectiveness of feature types and model architectures, examine practical implications, and consider the broader significance of benchmark-driven evaluation for learned cost models \cite{StatisticalLearning}.

\subsection{Performance Insights and Feature Contributions}

Our experiments demonstrate that accurate query runtime prediction is feasible using features derived directly from real execution plans in PostgreSQL \cite{PostgreSQLDatabase,ExplainAnalyze}. The refined XGBoost model outperformed all baselines with a mean squared error (MSE) of 0.3002 and ±10\% accuracy of 65.52\% \cite{XGBoost}, validating the core premise that learned models can improve upon traditional cost estimators \cite{QueryOptimizerSurvey}.

Importantly, we observe that feature design plays a critical role in model performance. Scalar features (e.g., estimated rows, startup cost, actual time) provide strong individual signals, but alone fail to capture contextual dependencies \cite{CardinalityEstimation}. Adding structural features—such as plan node depth, operator type, and subtree width enabled the model to understand hierarchical plan execution patterns \cite{DaliboVisualizer}. Semantic features derived from TF-IDF embeddings of the SQL text further improved generalization across queries with similar structure but different filtering predicates \cite{Jones1972,SentenceBERT}.

Feature importance analysis from XGBoost highlighted that actual row counts, estimated plan rows, and node depth were among the most influential attributes \cite{XGBoost,CardinalityEstimation}, reinforcing the value of combining multiple perspectives on plan behavior.

\subsection{Model Comparisons and Generalization}

The comparative analysis between learning approaches revealed clear trade-offs. Classical linear models underfit the data \cite{RegressionAnalysis}, and deep sequential models like LSTM struggled due to limited training samples and architectural constraints \cite{LSTM}. XGBoost, by contrast, provided a robust and interpretable alternative, handling non-linearities and heterogeneous feature types effectively without the high data requirements of neural networks \cite{XGBoost}.

The LSTM model underperformed across all evaluation metrics (MSE = 0.8915, R\textsuperscript{2} = 0.0216, ±10\% Accuracy = 17.24\%) \cite{LSTM}, highlighting the limitations of sequence models when plan trees are flattened into linear inputs and when model complexity exceeds dataset scale \cite{DeepCostModel}. These results suggest that tree-based models remain better suited for structured, low-volume data in query performance prediction tasks \cite{RandomForests}.

\subsection{Answering the Research Questions}

\textbf{RQ1: Can real execution traces from PostgreSQL be effectively leveraged to learn predictive models for total query runtime?}  
Yes. Execution traces from \texttt{EXPLAIN ANALYZE} contain rich operator-level supervision signals, including both estimated and actual metrics \cite{ExplainAnalyze,DepeszAnalyzer}. Our system successfully learned predictive models from these traces, and the XGBoost model demonstrated high accuracy and low error rates \cite{XGBoost}, confirming that real execution traces are valuable inputs for learning-based cost models.

\textbf{RQ2: How do different categories of features—scalar, structural, and semantic—contribute to the accuracy of learned cost models?}  
Each category of features added incremental value. Scalar metrics formed a strong baseline \cite{ExplainAnalyze}, structural features enabled the model to capture plan topology and execution flow \cite{DaliboVisualizer}, and semantic features improved contextual understanding of query behavior \cite{Jones1972,SentenceBERT}. The combination of all three yielded the best performance, underscoring the importance of multi-perspective feature integration.

\textbf{RQ3: Which types of ML models are most effective for learned runtime prediction in low-data regimes?}  
Tree-based ensemble models such as XGBoost provided the best trade-off between performance, robustness, and interpretability \cite{XGBoost,RandomForests}. Unlike deep models that suffer from overfitting or require extensive data and tuning \cite{LSTM}, XGBoost generalized well to unseen queries and provided insight into feature contributions, making it a practical choice in realistic scenarios with limited training data.

% \textbf{RQ4: Can benchmark-driven evaluation using TPC-H provide robust, repeatable insights into learned cost modeling approaches?}  
% Yes. The use of TPC-H allowed for controlled, systematic query generation with varied structures and selectivities \cite{TPCHBenchmark,TPCHSpec}. Our dataset—comprising over 1000 plan instances generated through parameterized query variation—enabled consistent model training and evaluation. TPC-H thus provided a reproducible foundation for benchmarking and comparing learned models in query performance prediction.

\textbf{RQ4: Can semantic understanding of the queries improve the accuracy of the prediction model?}  
Yes. Our experiments demonstrated that incorporating semantic features—specifically TF-IDF embeddings of the SQL query text \cite{Jones1972}—provided meaningful improvements in model performance. In particular, the refined XGBoost model that integrated these embeddings alongside scalar and structural plan features \cite{SentenceBERT,ExplainAnalyze,DaliboVisualizer} achieved the highest accuracy, with a mean squared error of 0.3002 and over 65\% of predictions falling within ±10\% of the actual runtime. This confirms that capturing query intent and complexity at the semantic level complements execution plan statistics and enhances the model's ability to generalize across diverse query templates.

\subsection{Practical Relevance}

From a deployment perspective, our findings suggest that integrating lightweight learned models—such as XGBoost \cite{XGBoost}—into cost estimation pipelines can offer meaningful improvements in query planning \cite{QueryOptimizerSurvey}. The ability to predict runtimes more accurately supports better resource allocation, job scheduling, and SLA adherence in database systems, particularly in multi-tenant or cloud-hosted environments \cite{query_lifecycle}. The low inference overhead and interpretability further strengthen the case for adoption in production systems.

\subsection{Lessons and Limitations}

The study also reveals important limitations and areas for future exploration. First, dataset scale remains a bottleneck for training deep architectures effectively \cite{LSTM}. Second, our flattened sequence encoding for plan trees did not preserve full hierarchical semantics, limiting LSTM performance \cite{DeepCostModel}. Third, certain low-level hardware features (e.g., I/O, memory buffers) were not captured due to PostgreSQL's trace granularity \cite{ExplainDocs}. Expanding trace collection and exploring tree-based neural architectures (e.g., Tree-LSTMs or GNNs) may address these gaps.

\subsection{Summary}

In conclusion, our work shows that real plan traces, carefully engineered features, and structured models offer a reliable and effective path to query runtime prediction \cite{PostgreSQLDatabase,ExplainAnalyze}. By answering the core research questions, we highlight both the feasibility and challenges of learned cost estimation and lay the groundwork for future systems that integrate machine learning into the core of the query optimizer \cite{QueryOptimizerSurvey}.

\section{Related Work}
\label{sec:related_work}

In this section, we review the foundational literature that informed the design, implementation, and refinement of our learned query runtime prediction system. The review focuses on three primary threads: learned cost models for large-scale query processing, deep learning-based sequential models for cost estimation, and broader trends in learning-augmented query optimizers. Our discussion is grounded in studies such as CLEO\cite{BigDataCost}, RAAL\cite{DeepCostModel}, and recent surveys that shaped both our baseline and refined modeling choices.

\subsection{Learned Cost Models}
Moving beyond cardinality, some efforts propose replacing handcrafted cost functions with learned models \cite{BigDataCost}. RAAL (Resource Aware Attentional LSTM)~\cite{DeepCostModel} applies a sequential LSTM with dual attention to Spark SQL plans, explicitly weighting both structural plan dependencies and dynamic resource features (e.g.\ executor memory, cores) to predict runtime under varying cluster configurations. Yan Li et al.~\cite{Li2024}  extend this line with a resource‑aware deep cost model that embeds query plan trees and extracted resource allocations into an adaptive attention framework, further demonstrating improved prediction stability as both cluster resources and underlying data change.  

\subsection{Execution‑Time Prediction from Plans}
Fewer studies focus on predicting actual runtime, which is affected by hardware variability, disk I/O, caching, and runtime behavior beyond logical plans. The RAAL\cite{DeepCostModel} model described above is a notable example that incorporates resource consumption patterns into a sequential model. Our work differs from RAAL in that we focus purely on PostgreSQL and use a hybrid feature set combining structural and semantic elements without assuming access to fine‑grained resource metrics.

\subsection{PostgreSQL‑Centric and Benchmark‑Based Studies}
Several recent studies leverage benchmark-driven evaluations on open-source systems like PostgreSQL \cite{QueryOptimizerSurvey}. These studies demonstrate the feasibility of reproducible, plan-based learning in real systems. Our approach builds on this foundation by integrating structural and semantic plan features and evaluating multiple models, including a reimplementation of RAAL-style sequential learners \cite{LSTM}.

\subsection{Surveys \& System Support}
Zhu et al.~\cite{Zhu2024} offer a comprehensive survey of learned query optimizers, categorizing ML techniques for cardinality, cost, and plan enumeration, and highlighting end‑to‑end learned optimizers. They also introduce a middleware that decouples ML model development from DBMS internals to simplify deployment. Our work complements theirs by delivering a full-stack, PostgreSQL‑focused pipeline that incorporates semantic embeddings and benchmarking on TPC-H, along with detailed comparisons between tree‑based and sequential models.

\subsection{Summary}
In contrast to prior work that often targets optimizer-internal cost proxies or assumes extensive profiling infrastructure, our framework focuses on practical runtime prediction from PostgreSQL's native execution traces \cite{ExplainAnalyze}. By combining scalar plan metrics, structural hierarchy, and SQL semantics \cite{Jones1972,SentenceBERT}, we present a unified, extensible pipeline for learned query runtime modeling.

\section{Conclusion and Future Work}
\label{sec:conclusion}

This paper presented a data-driven framework for predicting SQL query runtimes in PostgreSQL \cite{PostgreSQLDatabase} using features extracted from actual execution plans \cite{ExplainAnalyze}. Motivated by the limitations of traditional cost models \cite{QueryOptimizerSurvey} and guided by four targeted research questions (RQ1--RQ4), we developed an end-to-end system that automates query generation, plan parsing, feature engineering, and supervised model training using parameterized TPC-H queries \cite{TPCHBenchmark,TPCHSpec}.

Through systematic experimentation, we demonstrated that:

\begin{itemize}
    \item \textbf{RQ1:} Real execution traces can be effectively leveraged for training predictive models, providing fine-grained operator-level statistics that capture true runtime behavior \cite{ExplainAnalyze,DepeszAnalyzer}.
    \item \textbf{RQ2:} Scalar, structural, and semantic features jointly contribute to model accuracy, with structural plan features playing a particularly critical role \cite{DaliboVisualizer,SentenceBERT,Jones1972}.
    \item \textbf{RQ3:} Tree-based models such as XGBoost strike an effective balance between performance and interpretability in low-data regimes, outperforming deep models like LSTM \cite{XGBoost,LSTM}.
    \item \textbf{RQ4:} Incorporating semantic understanding of queries through transformer based embeddings of SQL text enhances model accuracy by capturing intent and structure beyond plan statistics.
\end{itemize}

Our findings validate the utility of plan- and query-level features for runtime prediction and demonstrate the practical feasibility of incorporating learned models into the cost estimation pipeline of PostgreSQL \cite{PostgresDesign}.

\subsection*{Future Work}
Several directions remain open for advancing this line of work:

\begin{itemize}
    \item \textbf{Neural Architectures:} Future models may explore attention-based transformers or graph neural networks to capture execution plan hierarchies more effectively than sequential models \cite{SentenceBERT}.
    \item \textbf{Multi-System Support:} Generalizing the framework to other open-source and commercial DBMSs could expand its applicability in diverse deployment environments \cite{query_lifecycle}.
    \item \textbf{Adaptive Learning:} Incorporating online or incremental learning techniques can allow the system to adapt to changing workloads and system states over time \cite{StatisticalLearning}.
    \item \textbf{Query Optimizer Integration:} Embedding learned cost models directly into the PostgreSQL optimizer could improve plan selection accuracy and enable re-optimization based on observed execution behavior \cite{QueryOptimizerSurvey}.
\end{itemize}

In conclusion, this study bridges the gap between traditional cost estimation and learned query performance modeling, offering a reproducible, extensible framework for integrating ML into query optimization workflows \cite{StatisticalLearning,QueryOptimizerSurvey}.

\bibliographystyle{ACM-Reference-Format}
\bibliography{sample-base}

\end{document}